\documentclass[twocolumn]{article}
\usepackage{graphicx}
\newcommand{\eqref}[1]{{(\ref{#1})}}
\begin{document}

\huge

Self-diffusion in granular gases: An impact of particles' roughness\\

\large

Anna Bodrova, Nikolai Brilliantov\\

\small
\textit{Department of Physics, Moscow State University, 119991 Moscow, Russia}\\
\textit{Department of Mathematics, University of Leicester, Leicester LE1 7RH, UK}\\






\small
An impact of particles'  roughness on the self-diffusion coefficient $D$ in granular gases is
investigated. For a simplified collision model where the normal, $\varepsilon$, and tangential, $\beta$,
restitution coefficients are assumed to be constant we develop an analytical theory for the diffusion
coefficient, which takes into account non-Maxwellain form of the velocity-angular velocity distribution
function. We perform molecular dynamics simulations for a gas in a homogeneous cooling state and study
the dependence of the self-diffusion coefficient on $\varepsilon$ and $\beta$. Our theoretical results
are in a good agreement with the simulation data.


\normalsize

\section{Introduction}
\label{intro} Among numerous contributions by Isaac Goldhirsch to the theory of granular fluids
\cite{Goldhirsch:2003} are his pioneering works   on the hydrodynamic of gas of particles with the
rotational degrees of freedom \cite{Goldhirsch_PRL:2005}. We wish to dedicate this article, addressed
to  granular gases  of rough particles, to the memory of Isaac Goldhirsch.

Granular fluids are systems composed of a large number of macroscopic particles which suffer dissipative
collisions. In many respects they behave like ordinary molecular fluids and may be described by  the
standard tools of kinetic theory and hydrodynamics \cite{Goldhirsch:2003,book,PoeschelBrilliantov:2003}.
Among numerous phenomena, common to molecular and granular fluids, are Brownian motion, diffusion and
self-diffusion \cite{Brey_JSP:1999,SD,Garzo:2002,Puglisi:2002,SDGreenCubo,physa,Puglisi_JSP:2010}; in
the latter case tracers are identical to surrounding particles. The self-diffusion coefficient $D$ has a
microscopic and macroscopic (thermodynamic) meaning, e.g. \cite{book,SDGreenCubo}: Microscopically, it
characterizes the dependence on time of the mean-square displacement of a grain
$\langle\vec{R}^{2}(t)\rangle$:
\begin{equation}
\langle\vec{R}^{2}(t)\rangle=6\int\limits_{0}^{t}D(t^{\prime})dt^{\prime}\, , \label{R2int}
\end{equation}
while macroscopically,  it relates the macroscopic flux of tracers  $\vec{J}_s \left(\vec{r}, t
\,\right)$ (the index {\em s} refers to tracers) to its concentration gradient $ \vec{\nabla} n_s \left(\vec{r}, t \,\right)$:
\begin{equation}
\label{eq:Js_nablans} \vec{J}_s \left(\vec{r},t\,\right)= - D(t)  \vec{\nabla} n_s
\left(\vec{r},t\,\right) \,.
\end{equation}
With the continuity equation
$\frac{\partial n_s \left(\vec{r},t\,\right) }{\partial t} + \vec{\nabla} \vec{J}_s \left(\vec{r},t
\,\right)= 0$
the local tracers density $n_s\left(\vec{r},t\,\right)$  obeys the diffusion
equation:
\begin{equation}
\label{difcanon} \frac{\partial n_s \left(\vec{r}\,\right) }{\partial t} = D \vec{\nabla}^2 n_s
\left(\vec{r}\,\right)\,.
\end{equation}
If the state of a granular gas is not stationary, like, e.g. a  homogeneous cooling state, the diffusion
coefficient generally depends on time.

In the previous studies the self-diffusion coefficient has been calculated only for smooth particles,
that is, for particles which do not exchange  upon collisions the energy of their rotational motion
\cite{Brey_JSP:1999,SD,Garzo:2002,SDGreenCubo,physa,Puglisi_JSP:2010}. The macroscopic nature of
granular particles, however,  implies  friction between their surfaces and hence the rotational degrees
of freedom are unavoidably involved in grains dynamics.

In the present paper  we analyze the impact of particles  roughness on the self-diffusion in granular
gases -- rarified systems, where the volume of the solid fraction is much smaller than the total volume.
We calculate the self-diffusion coefficient  as a function of normal and tangential restitution
coefficients, which we assume to be constant.  We develop an analytical theory for this kinetic
coefficient and perform molecular dynamics (MD) simulations. The rest of the article is organized as
follows. In Sect.~\ref{sec:2} we formulate the model, introduce the necessary notations and give the
detailed description of the theoretical approach. In Sect.~\ref{sec:3} we present the results of MD
simulations and compare the numerical data with the predictions of our theory. In the last
Sect.~\ref{sec:4} we summarize our findings.

\section{Calculation of the self-diffusion coefficient}
\label{sec:2}

\subsection{Collision rules}
The dissipative collisions of grains are characterized by two quantities, $\varepsilon$ and $\beta$ --
the normal and tangential restitution coefficients. These coefficients relate respectively  the normal
and tangential components of the relative velocity between surfaces of colliding grains,
$\vec{g}=\vec{v}_{12}+\frac{\sigma}{2}\left(\vec{e}
\times\left(\vec{\omega}_1+\vec{\omega}_2\right)\right)$, before (unprimed quantities) and after (primed
quantities) a collision \cite{book,Zippelius:2006}:
\begin{equation}
\label{eq:2}
\left(\vec{g}^{\, n}\right)^{ \prime} =-\varepsilon \vec{g}^{\, n} \qquad \quad \left(\vec{g}^{\,
t}\right)^{ \prime} =-\beta\vec{g}^{\, t} \, .
\end{equation}
Here $\vec{v}_{12} = \vec{v}_{1}-\vec{v}_{2}$ is the relative translational velocity of particles,
$\vec{\omega}_1$ and $\vec{\omega}_2$ are their rotational velocities, the unit vector $\vec{e}$ is
directed along the inter-center vector $\vec{r}_{12}$  at the collision instant and $\sigma$ is  the
diameter of particles. The normal and tangential components of $\vec{g}$ read respectively, $\vec{g}^{\,
n} = \left( \vec{g} \cdot \vec{e} \,\right) \vec{e}$ and $\vec{g}^{\, t} = \vec{g} - \vec{g}^{\, n}$.
For macroscopic particles the normal restitution coefficient $\varepsilon$ can vary from $0$ (completely
inelastic impact) to $1$ (completely elastic collision)\footnote{For nano-particles the normal
restitution coefficient can attain negative values \cite{SBHB:2010}.}, while the tangential coefficient
$\beta$ ranges from $-1$ (absolutely smooth particles) to $+1$ (absolutely rough particles). From the
conservation laws and Eqs.~(\ref{eq:2}) follow the after-collision velocities and angular velocities of
particles in terms of the pre-collision ones, e.g. \cite{book,Zippelius:2006,Temp}:
\begin{eqnarray}
\label{vprime}
\vec{v}_{1,2}^{\, \prime} = \vec{v}_{1,2}\mp\frac{1+\varepsilon}{2}\vec{g}^n \mp\eta\vec{g}^t \nonumber
\\
\label{omegaprime}
 \vec{\omega}_{1,2}^{\, \prime}  = \vec{\omega}_{1,2}+ \frac{2\eta}{q\sigma}\left[\vec{e}\times\vec{g}^t\right] \, .
\end{eqnarray}
Here $\eta \equiv \frac{q(1+\beta)}{2(1+q)}$ characterizes friction,
$q=4I/m\sigma^2$, with $I$ and $m$ being respectively the moment
of inertia and mass of particles. Although the normal and tangential restitution coefficients generally
depend on the translational and rotational velocities of colliding particles and an impact geometry,
e.g. \cite{book,Walton:1993,Zippelius:2006,Poesc_tan_PRE:2008,SBHB:2010}, here we assume
for simplicity that the restitution coefficients are constants.

\subsection{Boltzmann  equation}

To compute the self-diffusion coefficient in granular gases two main  approaches may be exploited --
Green-Kubo approach, based on the time correlation functions of a dynamical variable (a particle
velocity for the case of $D$) and Chapman-Enskog method, where the evolution of velocity distribution
function is analyzed. It may be shown that these two are equivalent for the case of constant normal restitution coefficient \cite{book,SDGreenCubo}; here we adopt the latter approach. First we consider the force-free
granular gas in a homogeneous cooling state. We start from the Boltzmann equation for the distribution
function of tracers $f_s(\vec{v}, \vec{\omega}, \vec{r},t)$,
\begin{equation}
\label{eq:collin_trac} \left(  \frac{\partial}{\partial t} +  \vec{v}_1 \cdot \vec{\nabla} \right)
f_s\left(\vec{v}_1,\vec{\omega}_1, \vec{r}, t\right) = g_2\left(\sigma\right)I(f,f_s) \, ,
\end{equation}
where $f(\vec{v},\vec{\omega}, \vec{r}, t) $ is the velocity distribution function of the gas particles,
 $g_2\left(\sigma\right)$ is the contact value of the pair correlation function, which accounts for the
increasing collision frequency due to the excluded volume of grains and $I(f,f_s)$ abbreviates the
collision integral \cite{book,SDGreenCubo,Santos_FF:2011}:
\begin{eqnarray}
\nonumber && I(f,f_s) \!=\!  \sigma^2 \!\int \! d \vec{v}_2 \! \int \! d \vec{\omega}_2 \! \int \! d \vec{e}
\Theta \left(-\vec{v}_{12} \cdot \vec{e} \, \right) \left| \vec{v}_{12} \cdot \vec{e} \,\right| \times \\
 &&    \Big[  \frac{1}{\varepsilon^2 \beta^2 }\! f_s\left(\vec{v}^{\,\prime\prime}_1,
\vec{\omega}^{\,\prime\prime}_1\right) f\left(\vec{v}^{\,\prime\prime}_2,
\vec{\omega}^{\,\prime\prime}_2\right) -
f_s\left(\vec{v}_1,\vec{\omega}_1\right)f\left(\vec{v}_2,\vec{\omega}_2\right)
\Big]   \nonumber\\
\label{eq:collin}
\end{eqnarray}
Here  $\vec{v}^{\,\prime\prime}_{1,2}$ and $\vec{\omega}^{\,\prime\prime}_{1,2}$ denote the
pre-collision velocities and angular velocities of particles for the {\em inverse}  collision, i.e., for
the collision which ends up with  $\vec{v}_{1,2}$ and $\vec{\omega}_{1,2}$, the factor
$\Theta\left(-\vec{v}_{12} \cdot \vec{e} \, \right)$ selects only approaching particles, and
$\left|\vec{v}_{12} \cdot \vec{e} \,\right|$ gives the length of the collision cylinder. We assume that
the concentration of tracers
\begin{equation}
\label{eq:n_s}
  n_s(\vec{r},t) \equiv \int d \vec{v}_1 \int d \vec{\omega}_1 f_s(\vec{v}_1, \vec{\omega}_1,  \vec{r},t)
\end{equation}
is much smaller than the number density of the gas particles $n(\vec{r},t)=n={\rm const}$, so that
tracers do not affect the velocity distribution function $f$ for the gas, which is the solution of the
Boltzmann equation
\begin{equation}
\label{eq:Boltz_gas}  \frac{\partial}{\partial t} f\left(\vec{v},\vec{\omega}, t\right) =
g_2\left(\sigma\right)I(f,f)
\end{equation}
for  the homogeneous cooling state. The collision integral in the last equation is given by
Eq.~\eqref{eq:collin} with the distribution function of tracers $f_s$ substituted by that for the gas
particles $f$.

Solving the Boltzmann equation \eqref{eq:collin_trac} for $f_s \left( \vec{v},\vec{\omega},\vec{r},t
\right)$, which is also known as Boltzmann-Lorentz equation, one can write the diffusion flux
\begin{equation}
\label{eq:J_s_def} \vec{J}_s \left( \vec{r}, t \right) = \int d\vec{v} \int d\vec{\omega}\, \vec{v}
f_s(\vec{v}, \vec{\omega},  \vec{r},t)  \, ,
\end{equation}
and then, using the macroscopic equation \eqref{eq:Js_nablans}, find the  diffusion coefficient as the
coefficient at the concentration gradient  $\vec{\nabla} n_s \left(\vec{r},t\,\right)$.

In a  homogeneous cooling state a density of the system remains uniform, while the translational $T(t)$ and rotational $R(t)$ granular temperatures, defined through the average energies of translational and rotational motion, respectively,
\begin{equation}
\label{eq:T_tr} \frac32 n T =  \int \! d\vec{v} d\vec{\omega}\, \frac{mv^2}{2} f,  \qquad  \frac32 n R=
\int d \!\vec{v}  d\vec{\omega}\, \frac{I \omega^2}{2} f
\end{equation}
decay with time as $\partial T /\partial t = -\zeta T$ and $\partial R/\partial t = -\zeta_{R} T$,
where $\zeta$ and $\zeta_R$ are the translational and rotational cooling rates, e.g.
\cite{book,Zippelius:2006,Corr,Santos_FF:2011}.

The distribution function $f\left(\vec{v},\vec{\omega},t\right)$ is close to the Maxwell distribution and the deviations from the Max-wellian are twofold.
Firstly, the translational and angular velocity distributions  slightly differ from the Gaussian
distribution, which is accounted by the expansion of $f$ in Sonine polynomials series with respect to
$\vec{v}^{\, 2}$ and $\vec{\omega}^{\, 2}$. Secondly, a linear velocity and spin of particles are
correlated; this is quantified using an expansion in Legendre polynomials series with respect to $\theta
= \widehat{\vec{v}\, \vec{\omega}}$ --  the angle between the velocity $\vec{v}$ and angular velocity
$\vec{\omega}$ of a particle. Using the reduced quantities $\vec{c} = \vec{v}/v_T$ and $\vec{w} = \vec{\omega }/\omega_T$
with $v_T  \equiv \sqrt{2T/m}$ and $\omega_T  \equiv \sqrt{2R/I}$, the distribution function reads
\begin{eqnarray}
\nonumber
f \left(\vec{v},\vec{\omega}, t\right) =  \frac{n}{\pi^2v_T^3 \, \omega_T^3} e^{-c^2 -w^2} \Big[ 1+a_{20}
S_{1/2}^{(2)}(c^2)+
 \\ 
 \nonumber
a_{02} S_{1/2}^{(2)}(w^2) + a_{11} S_{1/2}^{(1)}(c^2) S_{1/2}^{(1)}(w^2) \!+ \\ \bar{b}c^2w^2 P_2(\cos \theta) \Big] ,
\label{eq:disfunc}
\end{eqnarray}
where only  leading terms in the expansion are kept. The Sonine and Legendre polynomials in
Eq.~\eqref{eq:disfunc} are $S_{1/2}^{(1)}(x) = \frac32 -x$, $S_{1/2}^{(2)}(x) = \frac18(15-20x -4x^2)$ and $P_2(x) = \frac32 \left(x^2 -\frac13 \right)$.
The distribution function attains the scaling form \eqref{eq:disfunc} after some relaxation time. At
this stage, which will be addressed below, the coefficients $a_{20}$, $a_{11}$, $a_{02}$ and $\bar{b}$
are constants. These coefficients
have been analyzed  in detail in Refs. \cite{Santos_FF:2011,Corr}. They are complicated nonlinear functions of
$\varepsilon$, $\beta$ and $r=R/T$, which are too cumbersome to be given here.

The temperature ratio $r$ also attains a steady state value in the scaling regime and the cooling rates
become equal, $\zeta=\zeta_R$ \cite{Santos_FF:2011,Corr}. The steady-state value of $r=R/T$ may be
obtained as an appropriate root of the eighth-order algebraic equation \cite{Santos_FF:2011}. The
cooling rate $\zeta$ may be represented in the following way:
\begin{eqnarray}
\nonumber
&&\zeta =  -\frac{2}{3n T} \int d\vec{v}_1 \int d \vec{\omega}_1 \frac{mv_1^2}{2} I
(f,f) = 2\tau_{E}^{-1} \zeta^* \, , \\
&&\zeta^*=  \left( \frac14(1-\varepsilon^2) +\eta (1-\eta) \right) \left( 1
+\frac{3}{16}a_{20}  \right) - \nonumber \\
&&- \left( 1-\frac{a_{20}}{16} +\frac{a_{11}}{4} -
 \frac{\bar{b}}{8} \right) \frac{\eta^2}{q}r
 \label{eq:psi_tr}
\end{eqnarray}
and $\zeta_R$ may be obtained similarly \cite{Santos_FF:2011}. Here $\tau_{ E}$ is the Enskog relaxation
time
\begin{equation}
\label{eq:tau_E} \tau_{E}^{-1}(t)  = \frac83 g_2(\sigma) \sigma^2 n \sqrt{\frac{\pi T (t) }{m}}\, ,
\end{equation}

\subsection{Chapman-Enskog scheme}
To find $f_s$ we need to solve the Boltzmann-Lorentz equation \eqref{eq:collin_trac}. It may be done
approximately, using the Chapman-Enskog approach  based on two simplifying assumptions: (i)~$f_s
\left(\vec{v}, \vec{\omega},\vec{r},t \right)$ depends on space and time only trough the macroscopic
hydrodynamic fields and (ii)~$f_s \left( \vec{v},\vec{\omega},\vec{r},t \right)$ can be expanded in
terms of the field gradients. In the gradient expansion
\begin{equation}
\label{eq:ChEns_fs} f_s= f_s^{(0)} + \lambda f_s^{(1)} + \lambda^2 f_s^{(2)} \ldots \,
\end{equation}
a formal parameter $\lambda$ is introduced, which indicates the power of the field gradient; at the end
of the computations it is set to unity, $\lambda=1$.

In a homogeneous cooling state with a lack of macroscopic fluxes only two hydrodynamic fields are
relevant --  the number density of tracers, $n_s(\vec{r},t)$, and the total temperature $T_{\rm tot} =
T(t)+R(t)$. The former corresponds to conservation of particles, and the latter to conservation of
energy in the elastic limit. $T_{\rm tot}$ is equal for the gas and tagged particles due to their
mechanical identity. In the scaling regime, when the Chapman-Enskog approach is applicable, the
translational, rotational and total temperature are linearly related. Hence one can use any of these
fields in the Chapman-Enskog scheme and we choose $T(t)$ here  without the loss of  generality. Thus,
the two relevant hydrodynamic fields satisfy in the homogeneous cooling state the following  equations:
\begin{eqnarray}
\label{eq:dndt_dTdt_lamb}
\frac{\partial n_s}{\partial t}  = \left( \frac{\partial^{(0)}}{\partial t} + \lambda \frac{\partial^{(1)}}{\partial t} + \lambda^2 \frac{\partial^{(2)}}{\partial t} + \ldots \right) n_s
                                       =\lambda^2 D \nabla^2 n_s  \nonumber   \\ \\
 \frac{\partial T  }{\partial t}   = \left( \frac{\partial^{(0)}}{\partial t} + \lambda \frac{\partial^{(1)}}{\partial t} + \lambda^2 \frac{\partial^{(2)}}{\partial t} + \ldots \right) T
                                       = - \zeta T \, ,\nonumber
\end{eqnarray}
where $ \partial^{(k)} \psi  / \partial t $ indicates that only terms of $k$-th order with respect to
the field gradients are accounted.

Substituting Eq.~\eqref{eq:ChEns_fs} into the Boltzmann-Lorentz equation \eqref{eq:collin_trac} and
collecting terms of the same order of $\lambda$ (that is of the same order in gradients) we obtain
successive equations for $f_s^{(0)}(\vec{v},\vec{\omega},\vec{r},t)$, $ f_s^{(1)}
(\vec{v},\vec{\omega},\vec{r},t)$, etc. The zeroth-order equation in the gradients yields
$f_s^{(0)}(\vec{v},\vec{\omega},\vec{r},t)$ for the homogeneous cooling state. Since the tracers are
mechanically identical to the rest of the particles, $f_s^{(0)}(\vec{v},\vec{\omega},\vec{r},t)$ is
simply  proportional to the distribution function of the embedding gas:
\begin{equation}
\label{eq:fs_homog} f_s^{(0)}(\vec{v},\vec{\omega},\vec{r},t) = \frac{n_s \left(\vec{r}, t \right) }{n}
f(\vec{v},\vec{\omega},t) \,.
\end{equation}
The first-order equation then reads
\begin{equation}
\label{eq:Boltz1_tagg} \frac{\partial^{(0)} f_s^{(1)}}{\partial t}\! +\!\frac{\partial^{(1)}
f_s^{(0)}}{\partial t} + \vec{v}_1 \vec{\nabla} f_s^{(0)}\! =\! g_2(\sigma) I \left( f, f_s^{(1)}
\right) \!.
\end{equation}
Using Eqs. (\ref{eq:dndt_dTdt_lamb}), one can show, that the first term in the left-hand side of Eq.~\eqref{eq:Boltz1_tagg} is equal to
\begin{equation}
  \label{eq:dt0f_s1}
    \frac{\partial^{(0)} f_s^{(1)}}{\partial t}\! = \! \frac{\partial^{(0)} n_s}{\partial t}
    \frac{\partial f_s^{(1)} }{\partial n_s} \!+ \! \frac{\partial^{(0)} T}{\partial t}
    \frac{\partial f_s^{(1)} }{\partial T }\! = \! -\zeta T \frac{\partial f_s^{(1)}}{\partial T},
\end{equation}
since $\partial^{(0)} n_s/ \partial t $. Similarly, we obtain that $\partial^{(1)} f_s^{(0)}/\partial t
= 0$. Substituting Eq.~\eqref{eq:fs_homog} with the space independent $n$ and $f$ into the last term in
the left-hand side of Eq.~\eqref{eq:Boltz1_tagg}, we arrive finally at the equation for $f_s^{(1)}$:
\begin{equation}
\label{eq:forf1s} \zeta T \frac{\partial f_s^{(1)}}{\partial T} + g_2(\sigma) I \left( f, f_s^{(1)}
\right) = \frac{f}{n} \left( \vec{v}_1 \cdot \vec{\nabla} n_s \right)    \, .
\end{equation}
We search for the  solution of Eq.~\eqref{eq:forf1s} in the form
\begin{equation}
\label{eq:form_of_f1s}
 f_s^{(1)} = \vec{G} \left(\vec{v}_1,  \vec{\omega}_1, t \right) \cdot
 \vec{\nabla} n_s \left( \vec{r}, t \right) \,,
\end{equation}
which implies with Eq.~\eqref{eq:J_s_def} the diffusion flux
\begin{equation}
\label{eq:D_via_Gv} \vec{J}_s \!=\!\int  d \vec{v}_1  \int d \vec{\omega }_1 \, \vec{v}_1 \left( \vec{G}
 \cdot \vec{\nabla} n_s \right)
       \!=\!-D  \vec{\nabla} n_s \, ,
\end{equation}
where we take into account the isotropy of zero-order function $f_s^{(0)}=f_s^{(0)}(|\vec{v}\,|,|\vec{\omega}\,|)$. From
Eq.~\eqref{eq:D_via_Gv} we obtain (cf. \cite{book}):
\begin{equation}
\label{eq:D_viaGvfinal} D= - \frac13 \int d \vec{v}_1 \int d \vec{\omega }_1 \, \vec{v}_1 \cdot \vec{G}
\,.
\end{equation}
We substitute $f_s^{(1)}$ from Eq.~\eqref{eq:form_of_f1s} into  Eq.~\eqref{eq:forf1s} discarding the
factor $\vec{\nabla} n_s$. Then we multiply it by $\frac13 \vec{v}_1$ and integrate over $\vec{v}_1$ and
$\vec{\omega}_1$ to obtain
\begin{eqnarray}
\label{eq:Gv_int_overv}
\zeta T \frac{\partial}{\partial T} \frac13 \!\!\int \!d \vec{v}_1 \!\int
\!\!d \vec{\omega}_1  \vec{v}_1 \cdot \vec{G} + \frac{g_2(\sigma)}{3} \int \! \! d \vec{v}_1 \! \int
\!\!d \vec{\omega}_1  \times \nonumber \\ \times \vec{v}_1 \cdot I \left( f, \vec{G} \right) =
\frac{2}{3nm} \int \!d \vec{v}_1 \! \int \! d \vec{\omega}_1  \frac{mv_1^2}{2} f .
\end{eqnarray}
The first term in the left-hand side equals $ - \zeta T \partial D / \partial T$, while the right-hand
side equals $(T/m)$, according to Eq.~\eqref{eq:T_tr}. The structure of Eq.~\eqref{eq:forf1s} dictates
the  Ansatz\footnote{Only the first term of  the expansion $B=B_0+B_1 S_{3/2}^{(1)}(v^2/v_T^2)+B_2
S_{3/2}^{(2)}(v^2/v_T^2) + \ldots$ is used in Eq.~\eqref{eq:G_via_vfb0}, see e.g. \cite{ChapmanCowling}}
\begin{equation}
\label{eq:G_via_vfb0} \vec{G} \left(\vec{v}, \vec{\omega},t \right) \propto \vec{v} f\left(\vec{v},
\vec{\omega},t \right) = B_0 \vec{v} f \, ,
\end{equation}
where we assume that the mean value of the rotational velocity $\vec{\omega}$ is equal to zero. The unknown constant $B_0$ is to be determined from the above equation. With this Ansatz the
second term in the left-hand side of Eq.~\eqref{eq:Gv_int_overv} reads
\begin{eqnarray}
    \label{eq:int_vI_Gf1}
      &&\frac{g_2(\sigma)}{3}   \int \!d\vec{\omega}_1 d\vec{v}_1  \vec{v}_1  I\left(f, \vec{G} \right)
      = \frac{g_2(\sigma)\sigma^2 }{3} \frac{B_0}{2} \! \int \!d\vec{v}_1 d\vec{v}_2   \nonumber \\
      &&~~~~~~~~\times \int \!  d\vec{\omega}_1 \!
      d\vec{\omega}_2 \!\int \!d\vec{e} \, \Theta \left(-\vec{v}_{12} \cdot \vec{e} \, \right)
      \left|\vec{v}_{12} \cdot \vec{e} \, \right|
       \\
      &&~~~~~~~~ \times  f\left(\vec{v}_1,\vec{\omega}_1 \right)  \! f\left(\vec{v}_2,  \vec{\omega}_2\right)
      \left( \vec{v}_1 - \vec{v}_2 \right) \cdot \left( \vec{v}^{\,\prime}_1 - \vec{v}_1 \right)  .\nonumber
\end{eqnarray}
To evaluate the expression in
Eq.~\eqref{eq:int_vI_Gf1} we use the collision rule \eqref{vprime} for the factor $\left(
\vec{v}^{\,\prime}_1 - \vec{v}_1 \right)$ and Eq.~\eqref{eq:disfunc} for the distribution function $f$.
After a straightforward algebra we arrive at
\begin{equation}
\label{eq:vIG} \frac{g_2(\sigma)}{3}   \int \!d\vec{\omega}_1 \!\int \!d\vec{v}_1  \vec{v}_1 I\left(f,
\vec{G} \right)
      = -\frac{B_0 Tn}{m} \tau_{v,\,{\rm ad}}^{-1},
\end{equation}
where
\begin{equation}
\label{eq:tau_ad} \tau_{v,\,{\rm ad}}^{-1} = \left( \frac{1+\varepsilon}{2} + \eta  \right) \left(1
+\frac{3 }{16} a_{20} \right) \tau_{E}^{-1}\, .
\end{equation}
Note that $\tau_{v,\,{\rm ad}}$ has a physical meaning of the relaxation time of the velocity
correlation function of granular particles, e.g. \cite{book,SDGreenCubo}. Eq.~\eqref{eq:tau_ad} gives
the generalization of this quantity for the case of rough grains. If we express the unknown constant
$B_0$ using the relation
$$
   D= -\frac{B_0}{3} \int d \vec{v}_1 \int d \vec{\omega}_1\, \vec{v}_1 \cdot \vec{v}_1 f(\vec{v}_1,
   \vec{\omega}_1,t)
     = -  \frac{B_0Tn}{m}\, , \nonumber
$$
according to Eqs.~(\ref{eq:D_viaGvfinal},\, \ref{eq:T_tr}), we recast \eqref{eq:Gv_int_overv} into the
form
\begin{equation}
\label{eq:forD_final} -\zeta T \frac{\partial D}{\partial T} +  D \tau_{v,\,{\rm ad}}^{-1} = \frac{T}{m}
\,.
\end{equation}
As it follows from Eqs.~\eqref{eq:psi_tr}, \eqref{eq:tau_E} and \eqref{eq:tau_ad},   $\tau_{v,\,{\rm
ad}}^{-1} \propto \tau_{{E}}^{-1} \propto \sqrt{T}$ and $\zeta T \frac{\partial}{\partial T}  \propto
\zeta  \propto \tau_{{E}}^{-1} \propto \sqrt{T}$. At the same time, the right-hand side of
Eq.~\eqref{eq:forD_final} scales as $\propto T$, which implies that  $D \propto \sqrt{T}$. Therefore $T
\frac{\partial D}{\partial T} = \frac{D}{2}$ and we obtain the solution of Eq.~\eqref{eq:forD_final},
\begin{equation}
D(t)  = \frac{T}{m} \left[ \tau_{v, \,{\rm ad}}^{-1}(t) - \frac12 \zeta (t) \right]^{-1} \, .
\label{eq:D_byond_ad1}
\end{equation}
Substituting $\tau_{v,\,{\rm ad}}$ and $\zeta$ given by Eqs.~\eqref{eq:psi_tr} and \eqref{eq:tau_ad}
into Eq.~\eqref{eq:D_byond_ad1} we  finally find:
\begin{eqnarray}
\label{eq:D_rough}
&&D  = \\
&&\frac{D_{E}}{\left(1\!+\!\frac{3}{16}a_{20} \right)
\left(\frac{(1+\varepsilon)^2}{4} \!+\! \eta^2 \right) \!+\!\left(1 \!-\!\frac{a_{20}}{16}
\!+\!\frac{a_{11}}{4}  -\frac{\bar{b}}{8}\right)\frac{\eta^2}{q}r } \nonumber
\end{eqnarray}
where $D_{E}(t) \equiv \left(T(t)/m \right) \tau_{E}(t)$ is the Enskog value for the self-diffusion
coefficient for  given temperature $T(t)$. Note that for $\eta=0$, i.e., for smooth particles,
Eq.~\eqref{eq:D_rough} reproduces the previously known result, e.g. \cite{book,SDGreenCubo}.

All the expansion coefficients $a_{20}$, $a_{11}$, $a_{02}$ and $\bar{b}$ are  small and with a reasonable accuracy one can use the Maxwell approximation
for $r$ \cite{Temp,Santos_FF:2011}:
\begin{eqnarray}
\label{eq:r_steady} r &=& \sqrt{1+C^2}+C \\
C& \equiv & \frac{1+q}{2q(1+\beta)} \left[ \frac{1- \varepsilon^2}{1+\beta} (1+q) -(1-q)(1-\beta)
\right]  \nonumber
\end{eqnarray}
and for $D$, which simplifies in this case to
\begin{equation}
\label{eq:D_roughMax}
 D  = D_{E} \left( \frac{(1+\varepsilon)^2}{4} +\eta^2 \left( 1+ \frac{r}{q} \right) \right)^{-1} \,.
\end{equation}

It is rather straightforward to perform similar calculations for a driven granular gas in the
white-noise thermostat. The Boltzmann equation reads in this case:
\begin{equation}
\frac{\partial}{\partial t} f\left(\vec{v},\vec{\omega}, t\right)-\frac{\chi^2}{2}\left(\frac{\partial}{\partial\vec{v}_1}\right)^2 f =
g_2\left(\sigma\right)I(f,f)\,,
\end{equation}
where $\chi$ characterizes the strength of the stochastic force. The gas temperature and the coefficients $a_{20}$, $a_{11}$, $a_{22}$ are then determined by the intensity of the noise \cite{Santos_FF:2011}. The temperature rapidly relaxes to a constant value, therefore only one hydrodynamic field $n_s(r, t)$ is relevant in the Chapman-Enskog scheme. Performing then all steps as in the case of HCS, we finally arrive at the self-diffusion coefficient for a driven granular gas:
\begin{eqnarray}
\label{eq:D_termo}
D_{\rm w.n.th.}  = \frac{D_{E}}{\left(1 \!+\!\frac{3}{16}a_{20} \right)
\left(\frac{1+\varepsilon}{2} + \eta \right) }
\end{eqnarray}

\section{Molecular dynamics simulations}
\label{sec:3} To check the predictions of our theory we perform molecular dynamics (MD) simulations
\cite{CompBook} for a force-free system, using 8000 spherical particles of radius $\sigma/2=1$ and mass
$m=1$ in a box of length $L_{\rm box}=132$ with periodic boundary conditions. We confirmed that the
system remained in the HCS during its evolution. The initial translational and rotational temperatures
were equal $T(0)=R(0)=1$; the ratio between translational and rotational temperatures $r$ rapidly
relaxed to a constant value, indicating that the application of our theory, based on the scaling form of
the distribution function, is valid. To analyze the diffusion coefficient we used the re-scaled time
$\tau$
\begin{equation}
d\tau = dt/ \tau_c(t),  \quad \tau_c^{-1}(t)= 4 \sqrt{\pi} g_2(\sigma) \sigma^2 n \sqrt{T(t)/m}.
\nonumber
\end{equation}
In the re-scaled time the self-diffusion coefficient attains at $\tau \gg 1$  a constant value [see
Eq.~\eqref{R2int}], $ D = \langle R^2\left(\tau \right) \rangle/\left(6\tau\right) $.

\begin{figure}[htbp]
    \includegraphics[width=0.98\columnwidth]{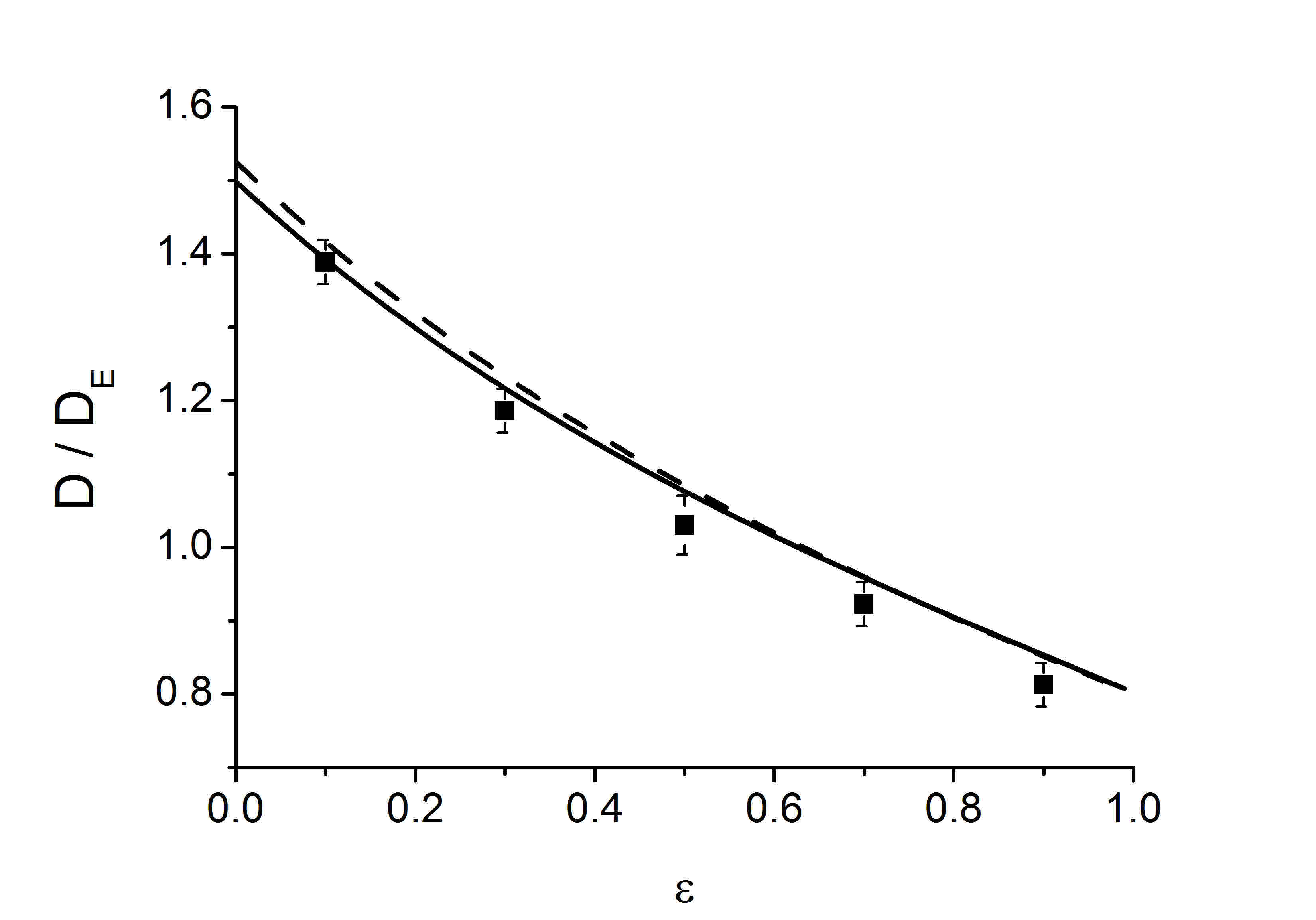}
    \includegraphics[width=0.98\columnwidth]{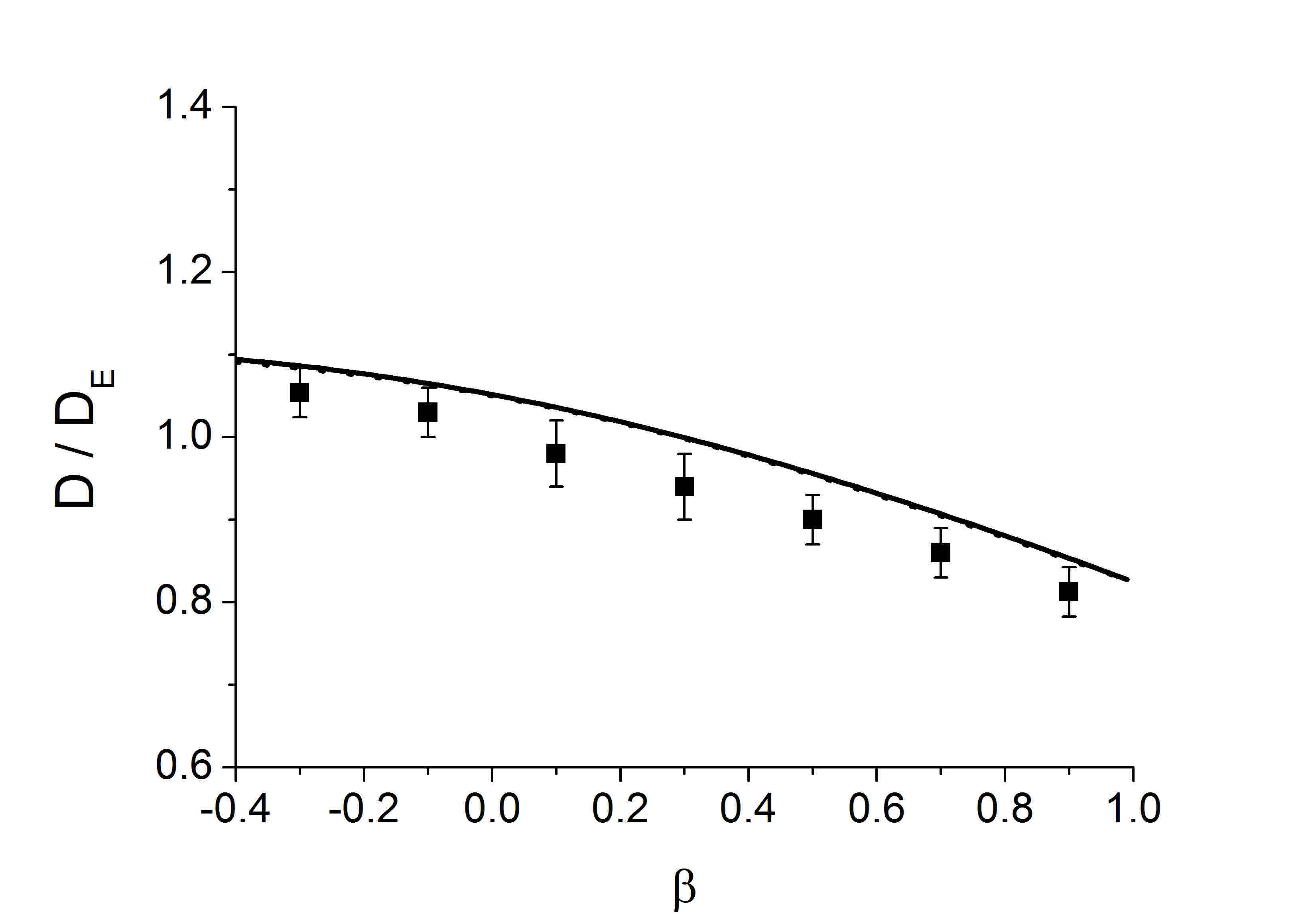}
    \caption
    {
    The dependence of the self-diffusion coefficient in a HCS on the normal $\varepsilon$ (upper panel) and tangential $\beta$ (lower panel) restitution coefficients.
    The MD simulation data (symbols) are  compared with the predictions of
    the theory for a gas in a HCS, Eq.~\eqref{eq:D_rough} (solid line) and Eq.~\eqref{eq:D_roughMax}
    (dashed line, Maxwellian approximation). The fixed restitution coefficients
    are: $\beta=0.9$ for the upper  and $\varepsilon =0.9$ for the lower panel.
    }
  \label{Gdv}
\end{figure}

The dependence of the self-diffusion coefficients on the normal and tangential restitution coefficients
is presented in Fig.~\ref{Gdv}, where the  molecular dynamics data are compared with the theoretical
predictions.
As it may be seen from the figure, the relative diffusion coefficient $D(t)/D_{E}(t)$ increases with
decreasing $\varepsilon$ -- the dependence, which has been already observed for smooth particles
\cite{Brey_JSP:1999,SD,Garzo:2002,book,SDGreenCubo,physa}. The physical nature of the effect is very
simple: With the increasing inelasticity, which suppresses the normal component of the after-collisional
relative velocity of particles, their trajectories become more stretched. This leads to the  increasing
correlation time and and hence, to a larger $D$. At the same time self-diffusion coefficient decreases
for large roughness ($\beta > -0.25$) with increasing tangential restitution coefficient, see
Fig.~\ref{Gdv}. It is not difficult to explain the observed behavior of $D$. Indeed, when the tangential
restitution coefficient $\beta$ increases from $-1$ (smooth particles) to $+1$ (absolutely rough
particles), the translational and rotational motion become more and more engaged [see
Eqs.~(\ref{vprime})] and the trajectories of particles more and more chaotic, that is, less stretched.
This eventually causes the decrease of the diffusion coefficient with roughness.

As it is follows from the Fig.~\ref{Gdv}, the theoretical predictions for the self-diffusion coefficient
are in a reasonably good agreement with the numerical data. The diffusion coefficient, calculated in the
framework of the Maxwellian approximation, practically does not differ from the full solution,
Eq.~\eqref{eq:D_roughMax}; the difference becomes apparent only for very high inelasticity
($\varepsilon<0.6$).  A slight overestimate by the theory of the value of $D$, obtained in molecular
dynamics,  may be possibly attributed to the simple Ansatz \eqref{eq:G_via_vfb0} for the first-order
function $f_s^{(1)}$, where only zero-order term of the expansion in the Sonine polynomials $
S_{3/2}^{(k)}(v^2/v_T^2) $ was used (see the footnote before Eq.~\eqref{eq:G_via_vfb0}). One probably
needs to extend the expansion and include the next order terms; this would be a subject of a future
study.

\section{Conclusion}
\label{sec:4}
 We have analyzed an impact of particles'  roughness on the self-diffusion coefficient in
granular gases. We use a simplified collision model with constant normal $\varepsilon$ and tangential
$\beta$ restitution coefficients. The former characterizes the collisional dissipation of  the normal
relative motion of particles at a collision, the latter -- of  the tangential one.  We develop an
analytical theory for the self-diffusion coefficient, taking  into account the deviation of the
velocity-angular velocity distribution function from the Maxwellian; we use the leading-order terms in
the expansion of this deviation in  Sonine and Legendre polynomials series. We notice  that the impact
of the non-Maxwellian distribution on the self-diffusion coefficient is small. To check the predictions
of our theory we perform the molecular dynamics simulations for a granular gas in a homogeneous cooling
state for different values of the normal and tangential restitution coefficients. We find that the
theoretical results are in a reasonably good agreement with the simulation data. Both the theory and
molecular dynamics demonstrate that the relative diffusion coefficient $D(t)/D_{E}(t)$ ($D_{E}$ is the
Enskog value of diffusion coefficient for smooth elastic particles) increases with decreasing normal restitution coefficient $\varepsilon$, similarly, as for a gas of smooth particles and decreases with increasing tangential restitution coefficient $\beta$ for large roughness ($\beta> -0.25$) .


\bibliographystyle{spmpsci}      

\end{document}